\documentclass[12pt]{article}
\begin{document}
      \sloppy

\def\AFOUR{%
\setlength{\textheight}{9.0in}%
\setlength{\textwidth}{5.75in}%
\setlength{\topmargin}{-0.375in}%
\hoffset=-.5in%
\renewcommand{\baselinestretch}{1.17}%
\setlength{\parskip}{6pt plus 2pt}%
}
\AFOUR
\def\car{\mathop{\square}}
\def\carre#1#2{\raise 2pt\hbox{$\scriptstyle #1$}\car_{#2}}

\parindent=0pt
\makeatletter
\def\section{\@startsection {section}{1}{\z@}{-3.5ex plus -1ex minus
   -.2ex}{2.3ex plus .2ex}{\large\bf}}
\def\subsection{\@startsection{subsection}{2}{\z@}{-3.25ex plus -1ex minus
   -.2ex}{1.5ex plus .2ex}{\normalsize\bf}}
\makeatother
\makeatletter
\@addtoreset{equation}{section}
\renewcommand{\theequation}{\thesection.\arabic{equation}}
\makeatother

\renewcommand{\a}{\alpha}
\renewcommand{\b}{\beta}
\newcommand{\g}{\gamma}           \newcommand{\G}{\Gamma}
\renewcommand{\d}{\delta}         \newcommand{\D}{\Delta}
\newcommand{\e}{\varepsilon}
\newcommand{\la}{\lambda}        \newcommand{\LA}{\Lambda}
\newcommand{\m}{\mu}
\newcommand{\n}{\nu}
\newcommand{\om}{\omega}         \newcommand{\OM}{\Omega}
\newcommand{\p}{\psi}             \newcommand{\PS}{\Psi}
\renewcommand{\r}{\rho}
\newcommand{\s}{\sigma}           \renewcommand{\S}{\Sigma}
\newcommand{\f}{{\phi}}           \newcommand{\F}{{\Phi}}
\newcommand{\vf}{{\varphi}}
\newcommand{\y}{{\upsilon}}       \newcommand{\Y}{{\Upsilon}}
\newcommand{\z}{\zeta}

\renewcommand{\AA}{{\cal A}}
\newcommand{\BB}{{\cal B}}
\newcommand{\CC}{{\cal C}}
\newcommand{\DD}{{\cal D}}
\newcommand{\EE}{{\cal E}}
\newcommand{\FF}{{\cal F}}
\newcommand{\GG}{{\cal G}}
\newcommand{\HH}{{\cal H}}
\newcommand{\II}{{\cal I}}
\newcommand{\JJ}{{\cal J}}
\newcommand{\KK}{{\cal K}}
\newcommand{\LL}{{\cal L}}
\newcommand{\MM}{{\cal M}}
\newcommand{\NN}{{\cal N}}
\newcommand{\OO}{{\cal O}}
\newcommand{\PP}{{\cal P}}
\newcommand{\QQ}{{\cal Q}}
\renewcommand{\SS}{{\cal S}}
\newcommand{\RR}{{\cal R}}
\newcommand{\TT}{{\cal T}}
\newcommand{\UU}{{\cal U}}
\newcommand{\VV}{{\cal V}}
\newcommand{\WW}{{\cal W}}
\newcommand{\XX}{{\cal X}}
\newcommand{\YY}{{\cal Y}}
\newcommand{\ZZ}{{\cal Z}}

\newcommand{\ch}{\widehat{C}}
\newcommand{\gh}{\widehat{\gamma}}
\newcommand{\W}{W_{i}}
\newcommand{\na}{\nabla}
\newcommand{\xint}{\dint d^4x\;}
\newcommand{\sla}{\raise.15ex\hbox{$/$}\kern -.57em}
\newcommand{\Sla}{\raise.15ex\hbox{$/$}\kern -.70em}
\def\h{\hbar}
\def\Lp{\displaystyle{\biggl(}}
\def\Rp{\displaystyle{\biggr)}}
\def\LP{\displaystyle{\Biggl(}}
\def\RP{\displaystyle{\Biggr)}}
\newcommand{\lp}{\left(}\newcommand{\rp}{\right)}
\newcommand{\lc}{\left[}\newcommand{\rc}{\right]}
\newcommand{\lac}{\left\{}\newcommand{\rac}{\right\}}
\newcommand{\identity}{\bf 1\hspace{-0.4em}1}
\newcommand{\complex}{{\kern .1em {\raise .47ex
\hbox {$\scriptscriptstyle |$}}
      \kern -.4em {\rm C}}}
\newcommand{\real}{{{\rm I} \kern -.19em {\rm R}}}
\newcommand{\rational}{{\kern .1em {\raise .47ex
\hbox{$\scripscriptstyle |$}}
      \kern -.35em {\rm Q}}}
\renewcommand{\natural}{{\vrule height 1.6ex width
.05em depth 0ex \kern -.35em {\rm N}}}
\newcommand{\tint}{\int d^4 \! x \, }
\newcommand{\intg}{\int d^D \! x \, }
\newcommand{\intm}{\int_\MM}
\newcommand{\tr}{{\rm {Tr} \,}}
\newcommand{\half}{\dfrac{1}{2}}
\newcommand{\pa}{\partial}
\newcommand{\pad}[2]{{\frac{\partial #1}{\partial #2}}}
\newcommand{\fud}[2]{{\frac{\delta #1}{\delta #2}}}
\newcommand{\dpad}[2]{{\displaystyle{\frac{\partial #1}{\partial
#2}}}}
\newcommand{\dfud}[2]{{\displaystyle{\frac{\delta #1}{\delta #2}}}}
\newcommand{\dfrac}[2]{{\displaystyle{\frac{#1}{#2}}}}
\newcommand{\dsum}[2]{\displaystyle{\sum_{#1}^{#2}}}
\newcommand{\dint}{\displaystyle{\int}}
\newcommand{\eg}{{\em e.g.,\ }}
\newcommand{\Eg}{{\em E.g.,\ }}
\newcommand{\ie}{{{\em i.e.},\ }}
\newcommand{\Ie}{{\em I.e.,\ }}
\newcommand{\nb}{\noindent{\bf N.B.}\ }
\newcommand{\etal}{{\em et al.}}
\newcommand{\etc}{{\em etc.\ }}
\newcommand{\via}{{\em via\ }}
\newcommand{\cf}{{\em cf.\ }}
\newcommand{\twiddle}{\lower.9ex\rlap{$\kern -.1em\scriptstyle\sim$}}
\newcommand{\qed}{\vrule height 1.2ex width 0.5em}
\newcommand{\grad}{\nabla}
\newcommand{\bra}[1]{\left\langle {#1}\right|}
\newcommand{\ket}[1]{\left| {#1}\right\rangle}
\newcommand{\vev}[1]{\left\langle {#1}\right\rangle}

\newcommand{\equ}[1]{(\ref{#1})}
\newcommand{\eq}{\begin{equation}}
\newcommand{\eqn}[1]{\label{#1}\end{equation}}
\newcommand{\eea}{\end{eqnarray}}
\newcommand{\eqa}{\begin{eqnarray}}
\newcommand{\eqan}[1]{\label{#1}\end{eqnarray}}
\newcommand{\ba}{\begin{array}}
\newcommand{\ea}{\end{array}}
\newcommand{\eqac}{\begin{equation}\begin{array}{rcl}}
\newcommand{\eqacn}[1]{\end{array}\label{#1}\end{equation}}
\newcommand{\qq}{&\qquad &}
\renewcommand{\=}{&=&} 
\newcommand{\cb}{{\bar c}}
\newcommand{\mn}{{\m\n}}
\newcommand{\pic}{$\spadesuit\spadesuit$}
\newcommand{\?}{{\bf ???}}
\newcommand{\Tr }{\mbox{Tr}\ }
\newcommand{\adot}{{\dot\alpha}}
\newcommand{\bdot}{{\dot\beta}}
\newcommand{\gdot}{{\dot\gamma}}

\global\parskip=4pt
\titlepage  \noindent
{
%
   \noindent

\hfill GEF-TH-7/2001 

\hfill LPTHE/01-22 \vspace{8mm}

\noindent
{\bf
{\Huge  Protected Operators in $\NN=2,4$}} \\

\noindent
{\bf {\Huge
Supersymmetric Theories  } }

\vspace{.5cm}
\hrule

\vspace{2cm}

\noindent
{\bf by
Nicola Maggiore$^{*}$
and
Alessandro Tanzini$^{**}$}

\noindent
{\footnotesize {\it
$^{*}$ Dipartimento di Fisica -- Universit\`a di Genova --
via Dodecaneso 33 -- I-16146 Genova -- Italy and INFN, Sezione di
Genova -- email: maggiore@ge.infn.it\\
$^{**}$LPTHE, Universit\'e Pierre et Marie Curie (Paris VI) et Denis Diderot
(Paris VII), Tour 16, 1er. \'etage, 4, Place Jussieu, 75252 Paris cedex 05,
France\\
-- email: tanzini@lpthe.jussieu.fr
} }

\vspace{2cm}
\noindent
{\tt Abstract~:}
The anomalous dimension of single and multi-trace composite operators
of scalar fields is shown to vanish at all orders of the perturbative series.
The proof hold for theories with $\NN=2$ supersymmetry with any number
of hypermultiplets in a generic representation of the gauge group.
It then applies to the finite $\NN=4$ theory as well as to non conformal
$\NN=2$ models.

\vfill\noindent
{\footnotesize {\tt Keywords:}
   Extended Supersymmetry, BRS quantization, AdS-CFT Correspondence
}
\newpage
\begin{small}
\end{small}

\setcounter{footnote}{0}


\section{Introduction}

It has been known for a long time that quantum field theories with
extended supersymmetry display a set of remarkable nonrenormalization
properties. The $\beta$--function of the gauge coupling was argued
to vanish to all orders of perturbation theory for the
${\cal N}=4$ Super Yang--Mills theory (SYM)~\cite{Sohnius:1981sn} and to
receive
only one--loop corrections for the ${\cal N}=2$ case~\cite{Howe:1983wj}.
The conjectured AdS/CFT correspondence~\cite{Maldacena:1998re}
has renewed the interest on the finiteness properties of these theories;
in fact, many of the tests of the correspondence have relied
on nonrenormalization properties, which are crucial in order to
ensure a meaningful comparison between the strong coupling regime,
accessible by type IIB supergravity computations, and the weak coupling
one, where the usual field theory techniques are reliable.

In particular, the prototype example of the
correspondence~\cite{Maldacena:1998re}
establish a duality between type IIB superstring on $AdS_5\times S^5$ and
$\NN=4$ SYM in the superconformal phase. In this context, it has been shown
that chiral gauge invariant operators belonging to short multiplets of the
superconformal group $SU(2,2|4)$ have vanishing anomalous
dimensions~\cite{Ferrara:1998pr}.
The analysis of correlation functions of
Chiral Primary Operators (CPO), which are the lowest scalar components
of such short superconformal multiplets, provided a
highly nontrivial check for the AdS/CFT
correspondence~\cite{Bianchi:1999ie,Eden:1999gh,Penati}.
Moreover, it recently led to discover new double-trace
operators which are protected
despite they do not obey any of the known shortening
conditions~\cite{Arutyunov:2001mh,Bianchi:2001cm}.

We remark that the nonrenormalization properties of
some CPO's does not necessarily rely on superconformal algebra, and it can
actually be shown also for theories with less number
of supersymmetries. This is indeed the case for
the gauge--invariant operator ${\rm Tr}\phi^2$, where
in the ${\cal N}=2$ formalism $\phi$ corresponds to the complex scalar
field of the vector multiplet. The anomalous dimension of this
operator has been shown to vanish to all orders of the perturbative
series in pure ${\cal N}=2$ SYM~\cite{Blasi:2000qw}; this allowed
to provide a rigorous proof of the celebrated nonrenormalization
theorems for the beta functions of ${\cal N}=2$ and ${\cal N}=4$
theories~\cite{Blasi:2000qw,Lemes:2001db}.

In this work we extend the analysis of~\cite{Blasi:2000qw} to higher rank
gauge--invariant polynomials of the scalar field $\phi$ and $\overline\phi$,
including both single and multi-trace operators.
It is worth to underline that our proof holds for ${\cal N}=2$ theories
with any number of hypermultiplets in a generic representation of the
gauge group. It then applies to the finite ${\cal N}=4$ theory as well
as to nonconformal models, which are at present object of intensive
research in extensions of the AdS/CFT duality to
phenomenologically more interesting gauge theories with a nontrivial
renormalization group flow~\cite{Polchinski:2001mx,Evans:2000ct}.

A geometrical interpretation of such nonrenormalization properties
can be given by resorting to the topologically twisted
formulation of the $\NN=2$ theory. In fact, in this context, the operators
we consider can be written as components of the Chern classes of the
universal bundle defined in \cite{Baulieu:1988xs}.

This work is organized as follows. In Section 2 we sketch the basic
features of the twisting procedure. In Section 3 we give the proof of
finiteness of the operators $\Tr \phi^{k},\ k\geq 2$ (single--traced) and
$\Pi_i\Tr \phi^{k_i},\ \Sigma_i k_i = k$ (multi--traced).
Our conclusions are drawn in Section 4, while we confined into
appendices the unavoidable technicalities of our proofs and the
explicit examples~$k=2,3,4$.

\section{The Twist}

It is well known that ${\cal N}=2$ SYM theory can be given a
``twisted'' formulation~\cite{Witten:1988ze}, which on flat
manifolds is completely equivalent to the conventional one.
Nonetheless, the use of the twisted variables
makes more evident and easily understandable
some important features, like the topological
nature of a subset of observables~\cite{Witten:1988ze}
and the nonrenormalization theorem concerning the beta function of the
gauge coupling constant~\cite{Blasi:2000qw}. In view of these advantages, we
choose to work in the twisted version of $\NN=2$ SYM theory.
We address the interested reader to the existing
literature~\cite{Witten:1988ze,Marino:1996sd}
for what concerns the many and deep aspects of the
twisting procedure. In this Section, we prefer rather to set up the
field theoretical background in which we shall work.

The usual $\NN=2$
supercharges $(\QQ^{i}_{\a},\overline{\QQ}_{j\adot})$ are
characterized by an index $i=1,2$, which counts the number of
supersymmetries, and a Weyl spinor index $\alpha$, which runs on the
same values as $i$~: $\alpha=1,2$. At the origin of the twist, is
Witten's idea of identifying the indices
\eq
i \equiv \a \ .
\eqn{1}
The resulting twisted supercharges $\QQ^{\b}_{\a}$ and
$\overline{\QQ}_{\a\adot}$ can then be rearranged, with usual
conventions~\cite{Bagger:1990qh}, into a scalar
$\delta$, a vector $\delta_{\mu}$ and a selfdual tensor
$\delta_{\mu\nu}$~:
\eqa
\delta &\equiv& \frac{1}{\sqrt{2}} \varepsilon^{\a\b}\QQ_{\a\b}
\label{delta} \\
\delta_{\m} &\equiv& \frac{1}{\sqrt{2}}
                         \overline\QQ_{\a\adot}(\s^{\m})^{\adot\a}
     \label{deltamu}                   \\
\delta_{\m\n} &\equiv& \frac{1}{\sqrt{2}}
(\s_{\m\n})^{\a\b}\QQ_{\b\a}\ .\label{deltamunu}
\eqan{a1}

The $\NN=2$ supersymmetry algebra, written for the twisted
supercharges, contains a subalgebra, formed by $\delta$,
$\delta_{\mu\nu}$ and by the translations $\partial_{\mu}$~:
\eqa
\d^{2} &=& 0 \nonumber \\
\{\d,\d_{\m}\} &=& \partial_{\m} \label{subalgebra}\\
\{\d_{\m},\d_{\n}\} &=& 0\ . \nonumber
\eqan{3}
We call the subalgebra~\equ{subalgebra} ``topological'', since it appears to
be common to all known topological field
theories~\cite{Birmingham:1991rh}, both of the
Witten and Schwartz type~\cite{Birmingham:1991ty}.
The three twisted supercharges
$\delta_{\mu\nu}$ turn out to be redundant, since they do not play any
role either in the classical definition or in the quantum extension of
the theory~\cite{Fucito:1997xm}.
For this reason, we shall discard them throughout
the rest of our reasoning.

Analogously to supercharges, the fields of
the $\NN=2$ gauge multiplet
$(A_{\m},\psi^{i}_{\a},\overline{\psi}^{i}_{\adot},\phi,\overline\phi)$,
which belong to the adjoint representation of the gauge group~$G$,
twist to
\eq
(A_{\m},\psi_{\m},\chi_{\m\n},\eta,\phi,\overline\phi)\ ,
\eqn{4}
where
\eqa
\eta &\equiv& \varepsilon^{\a\b}\psi_{[\a\b]}
\nonumber\\
\psi_{\m} &\equiv&
(\overline\s_{\m})^{\a\adot}\overline\psi_{\a\adot}
                         \label{twistedfields}\\
\chi_{\m\n} &\equiv&
(\s_{\m\n})^{\a\b}\psi_{(\a\b)}\nonumber
\eqan{5}
are the anticommuting fields resulting from the twist of the spinor
fields $(\psi^{i}_{\alpha},\bar\psi^{i}_{\adot})$ into
$(\psi_{\a\b},\overline\psi_{\a\adot})$. Notice that
the scalar fields $(\phi,\overline\phi)$ and the gauge boson $A_{\mu}$
are not altered by the twisting operation.

The $\NN=2$ SYM pure gauge action twists to the TYM action
\eq
S^{\NN=2}_{SYM}
(A_{\m},\psi^{i}_{\a},\overline{\psi}^{i}_{\adot},\phi,\overline\phi)
\longrightarrow
S_{TYM}
(A_{\m},\psi_{\m},\chi_{\m\n},\eta,\phi,\overline\phi)\ ,
\eqn{6}
which, in the $4D$ flat euclidean spacetime
reads~\cite{Fucito:1997xm}~:
\begin{eqnarray}
S_{TYM} &=&\frac 1{g^2}\Tr\displaystyle\int d^{4}x \left(
\frac
12F_{\mu \nu }^{+}F^{+\mu \nu }\;-\chi ^{\mu \nu }(D_\mu \psi _\nu
-D_\nu
\psi _\mu )^{+}\;\right.  \nonumber \\
&&\;\;\;\;\;\;\;\;\;\;\;\;\;\;\;\;\;+\eta D_\mu \psi ^\mu \;-\frac 12
\overline{\phi }D_\mu D^\mu \phi \;+\frac 12\overline{\phi }\left\{
\psi
^\mu ,\psi _\mu \right\}  \label{tym} \\
&&\;\;\;\;\;\;\;\;\;\;\;\;\left. \;\;\;-\frac 12\phi \left\{ \chi
^{\mu \nu},\chi _{\mu \nu }\right\} \;-\frac 18\left[ \phi ,
\eta \right] \eta
-\frac
1{32}\left[ \phi ,\overline{\phi }\right] \left[ \phi ,\overline{\phi}
\right] \right) \;,  \nonumber
\label{stym}\end{eqnarray}
where
$F_{\mu \nu }^{+} = F_{\mu \nu }+\frac 12\varepsilon _{\mu \nu \rho
\sigma}F^{\rho \sigma}$,
$\widetilde{F}_{\mu \nu }^{+}
=F_{\mu \nu }^{+}=
\frac 12
\varepsilon _{\mu \nu \rho \sigma }F^{+\rho \sigma }$
and \\
$(D_\mu \psi _\nu -D_\nu \psi _\mu )^{+}=(D_\mu \psi _\nu -D_\nu
\psi _\mu
)+\frac 12\varepsilon _{\mu \nu \rho \sigma }(D^\rho \psi ^\sigma
-D^\sigma
\psi ^\rho ).$

The symmetries of the action~\equ{tym} are the usual gauge invariance
      \eq
      \delta_{gauge}\ S_{TYM} =0 \ ,
      \eqn{gauge}
and the relevant twisted supersymmetries
      \eq
      \delta S_{TYM} = \delta_{\mu} S_{TYM} =0\ .
      \eqn{susy}
It is apparent from~\equ{twistedfields}, that the $4D$ flat euclidean twist
simply corresponds to a linear change of variables, therefore it is
not felt by the partition function
\eq
\ZZ = \int \DD\varphi\ e^{-S[\varphi]}\ .
\eqn{pathintegral}
Moreover, the stress-energy tensor of the theory is modified by the
twist only by a total derivative term which do not affect the translation
generators \cite{Marino:1996sd}.
This observation underlies the equivalence of the two,
twist--related, theories $\NN=2$ SYM and TYM when formulated on flat manifolds.
The theory has been
extended to the quantum level both in its
untwisted~\cite{Maggiore:1995dw} and
twisted~\cite{Fucito:1997xm} formulation.
Moreover, in~\cite{Blasi:2000qw} has been
demonstrated that a renormalization scheme does exist, in which the
$\beta$--function of the gauge coupling constant~$g$ receives one
loop contributions only.

\section{Protected Gauge Invariant Operators}

Let us  analyze the class of composite operators of the type
$\Tr \phi^{k},\ k\geq 2$ and $\Pi_i\Tr \phi^{k_i},\ \Sigma_i k_i = k$.
The aim of this Section is to show that
these operators are protected, in the sense
that are perturbatively finite, or, equivalently stated, they have
vanishing anomalous dimensions.

In order to prove this statement, let us consider the operator $\QQ$
defined as
\eq
\QQ = s +\om\d\ ,
\eqn{defQ}
where $s$ is the nilpotent BRS operator, $\delta$ is the twisted
scalar supercharge~\equ{delta}, and $\om$ is a global commuting ghost.

The operator $\QQ$ describes an invariance of the action $S_{TYM}$
\eq
\QQ S_{TYM} =0\ ,
\eqn{a2}
and it is nilpotent
\eq
\QQ^{2} =0\ ,
\eqn{a3}
since $s^{2}=\d^{2}=\{s,\d\}=0$.

For our purposes,  the $\QQ$--variations of
the scalar field $\phi$ and of the Faddeev--Popov ghost $c$ are
sufficient
\eqa
\mathcal{Q}\phi^{a} &=& f^{abc} c^{b}\phi^{c}
\nonumber \\
\mathcal{Q}c^{a} &=&\frac 12 f^{abc}c^{b}c^{c} - \om^{2}\phi^{a} \ ,
\eqan{qvariax}
where the index $a=1,\ldots,\mbox{dim}\ G$ counts the adjoint
representation matrices of the gauge group~$G$, and~$f^{abc}$ are the
structure constants of the corresponding Lie algebra.

The operators $\Tr \phi^{k}$ and $\Pi_i\Tr \phi^{k_i},\ \Sigma_i k_i = k$
are the observables of TYM (hence
$\NN=2$ SYM) theory~\cite{Witten:1988ze,Fucito:1997xm}; we can write them
as
\eq
D^{a_{1}.. a_{k}} \phi^{a_{1}}.. \phi^{a_{k}}\ ,
\eqn{a4}
where the $D^{a_{1}..a_{k}}$ are completely symmetric invariant
tensors of rank $k$
\begin{center}
      \begin{tabular}{llcl}
       $k=2$ &   $D^{a_{1}a_{2}}$      &=& $\delta^{a_{1}a_{2}}$ \ ; \\
       $k=3$ &   $D^{a_{1}a_{2}a_{3}}$ &=& $d^{a_{1}a_{2}a_{3}}$ \ ; \\
      &&\vdots&
\end{tabular}\end{center}
In particular, the $k=2$ operator $\Tr\phi^{2}$ can be thought to as
a kind of prepotential of the theory, since the whole action can be
written as
\eq
\S \approx -\frac{1}{3g^{2}}
\e^{\m\n\r\s}\d_{\m}\d_{\n}\d_{\r}\d_{\s} \int d^4x\;
\Tr \frac{\phi^{2}}{2}
\ ,
\eqn{action}
where this relation holds modulo BRS trivial
cocycles. The
equation~\equ{action} reflects the fact that $\Tr\phi^{2}$ contains
all the physical information of the theory, as it constitutes its inner
bulk (for a detailed discussion of this point,
see~\cite{Fucito:1997xm,Blasi:2000qw}).

A crucial point is that, in the space of local field polynomials
which are not necessarily analytic in the constant parameter~$\om$,
the operators~(\ref{a4}) can be written as $\QQ$--variations,
$i.e.$ in this space the cohomology of~$\QQ$ is empty. We stress that
quantum field theory rules require that the 1PI generating
functional~$\G$ must be an analytic function in $\om$ as well as in any
other parameter of the theory, whereas this condition can obviously be
relaxed for quantum insertions, like for instance~(\ref{a4}),
which can be rendered analytic by a multiplication for a suitable
power of~$\om$, as we shall see.

The following result holds (see proof in Appendix A )
\eqa
D^{a_{1}.. a_{k}} \phi^{a_{1}}.. \phi^{a_{k}} &=&
\QQ \left[D^{a_{1}.. a_{k}}
\left (
f_{0}^{(k)} c^{a_{1}}sc^{a_{2}}.. sc^{a_{k}}
+ f_{1}^{(k)} c^{a_{1}}sc^{a_{2}}.. sc^{a_{k-1}}\phi^{a_{k}}
\right.\right. \nonumber \\ && \left.\left.
\ \ \ \ \ \ \ \ \ \ \ \ \ + \ldots +
f_{k-1}^{(k)} c^{a_{1}}\phi^{a_{2}}.. \phi^{a_{k}}
\right )\right] \label{variax}\\
&=& \QQ\ \left[ D^{a_{1}.. a_{k}} \sum_{p=0}^{k-1}
f_{p}^{(k)} c^{a_{1}}sc^{a_{2}}.. sc^{a_{k-p}}
\phi^{a_{k-p+1}}.. \phi^{a_{k}}\right]\ ,\nonumber
\eqan{variaz}
or, in shorthand notation,
\eq
\Phi^{(k)}(\phi) = \QQ\ \PP^{(k)}(c,\phi)\ .
\eqn{shortvariax}
The coefficients $f_{p}^{(k)}$ are given by
\eq
f_{p}^{(k)} =
\frac{(-1)^{k-p}}{\om^{2(k-p)}}
\frac{(k-1)!k!}{p!(2k-p-1)!}\ \ ,\ \ 0\leq p\leq k-1\ ,
\eqn{a5}
and $sc^{a}$ is the ordinary BRS variation of the Faddeev--Popov
ghost $c$
\eq
sc^{a}=\left.\QQ c^{a}\right|_{\om=0}
=\half f^{abc}c^{b}c^{c}\ .
\eqn{a6}
Notice that the coefficient of the last term on the r.h.s.
of~\equ{variax}
is universal
\eq
f^{(k)}_{k-1} = -\frac{1}{\om^{2}}\ .
\eqn{a7}
It is worth to remark that
the operators (\ref{a4}) we are considering have an interesting
geometrical interpretation \cite{Baulieu:1988xs}.
Let us consider for the moment $k=2$;
by redefining the scalar field as $\phi \rightarrow \omega^2\phi$,
(\ref{shortvariax})
can be written as
\eq
\Tr \phi^2 = \QQ \Tr \Bigl(c\QQ c - {2\over 3}ccc\Bigr) \ .
\eqn{phi}
If we define the universal connection $\hat{A} = A + c$,
(\ref{phi}) can be seen as the $(p=0,q=4)$ component of the equation
\eqa
\Tr (\hat{F}\hat{F}) &=& \hat{d} \AA_{CS} \nonumber\\
&=& \hat{d} \Tr \Bigl(\hat{A}\hat{d}\hat{A} -
{2\over 3}\hat{A}\hat{A}\hat{A}\Bigr) \ ,
\eqan{uni}
where by $(p,q)$ we indicate the grading of the forms with respect to the
extended exterior derivative $\hat{d}=d+\QQ$, $p$ being the space-time form
degree and $q$ the ghost number.
Thus $\PP^{(2)}(c,\phi)$ can be understood as the $(0,3)$ component
of
the Chern-Simons form $\AA_{CS}$ of the universal connection $\hat{A}$, while
$\phi$ as the $(0,2)$ component of the corresponding curvature
$\hat{F}$.
Analogously, higher rank single-trace operators (\ref{a4}) can be
expressed in terms of $(0,k)$ components of higher Chern classes $\Tr
\hat{F}^k$, while multi-trace operators are products of 
Chern classes.

We now come to the proof of finiteness;
what is actually important in this sense is that
operators (\ref{a4}) can be expressed as the $\QQ$--variations of
polynomials in the fields $\phi$ and $c$, and, in particular, that
they are related to
$D^{a_{1}.. a_{k}} c^{a_{1}}sc^{a_{2}}.. sc^{a_{k}}$
(the first term on the r.h.s.
of~\equ{variax}, which belongs to
the cohomology of the ordinary BRS operator, and is known to be
finite to all orders of perturbation theory~\cite{Piguet:1992yg}.
It is then
natural to expect that the nonrenormalization property of the BRS
invariant odd polynomials in the ghost $c$ translates to (\ref{a4})
as well. As we shall see, this is indeed the case.

Let us consider now the classical action
\eq
\S = S_{TYM} + S_{gf} + S_{ext} + S_k\ .
\eqn{a8}
$S_{TYM}$ is the twisted $\NN=2$ SYM action~\equ{tym}, $S_{gf}$ is
the (Landau) gauge fixed term
\eq
S_{gf} = \QQ\ \Tr\, \tint\bar{c}\,\partial^{\mu}A_{\mu}\ ,
\eqn{a9}
$\bar{c}$ being the Faddeev--Popov antighost. According to the usual
BRS procedure to implement at the quantum level classical field
theories, in $S_{ext}$ the  nonlinear $\QQ$--variations of  the
fields~$\varphi$   are coupled to external fields~$\varphi^{*}$, sometimes
called ``antifields''
\eq
S_{ext} = \sum_{\varphi}\Tr\tint \varphi^{*}\QQ\varphi\ \ ,\ \
\varphi= A_{\mu},\chi_{\mu\nu},\psi_{\mu},\eta,\phi,\overline\phi,c\ ,
\eqn{a10}
and finally $S_k$ is given by
\eq
S_k = \om^{2k} \tint \left (
\a_{0} \Phi^{(k)} +
\a_{2k-1} \PP^{(k)} +
\sum_{p=2}^{2k-2} \a^{a_{1}..a_{p-1}}_{2k-p}
\partial_{a_{1}}..\partial_{a_{p-1}}\PP^{(k)}
\right)\ ,
\eqn{sk}
In the above expression for $S_k$,
the $\a$'s are external sources coupled
to~\mbox{$\Phi^{(k)}(\phi)$, $\PP^{(k)}(c,\phi)$} (defined
by~\equ{shortvariax}),
and its derivatives with respect to the ghost~$c$ ($\partial_{a}
\equiv \frac{\partial}{\partial c^{a}}$). They are antisymmetric in
their color indices $a$'s, and  are commuting or anticommuting objects
depending from the number
of ghosts~$c$ contained in the composite operators they are coupled to.
Such a number is represented by the lower indices, hence their
statistics is given by $(-1)^{2k-p}$. Notice that, thanks to the
overall multiplying factor~$\om^{2k}$, the functional $S_k$ is,
as it should for being part of the action, an analytic expression in~$\om$.

The whole action $\S$ satisfies the Slavnov--Taylor (ST) identity
\eq
\SS(\S) =\SS^{{old}}(\S) + \SS^{(k)}\S = 0\ ,
\eqn{stidentity}
where the usual ST operator
\eq
\SS^{{old}}(\S)=\sum_{\varphi} \Tr\tint \left (
\fud{\S}{\varphi} \fud{\S}{\varphi^{*}} +\bar{c}\fud{\S}{b} \right)
\eqn{a11}
has been implemented to the $\a$--sector as follows (see Appendix)
\eq
\SS^{(k)}\S = \tint \left (
\a_{2k-1} \fud{}{\a_{0}} +
\sum_{p=2}^{2k-3} \frac{p(1-p)}{2} f^{mna_{p-1}}
\a_{2k-p-1}^{mna_{1}..a_{p-2}} \fud{}{\a_{2k-p}^{a_{1}..a_{p-1}}}
\right) \S\ .
\eqn{a12}
Notice that the $\a$--sources, transforming one into the other, enter
the ST identity~\equ{stidentity} as BRS doublets,
which have the nice property
of keeping unaltered the cohomology of the (linearized) ST operator.

For the proof of finiteness, it is crucial that
an additional constraint on the theory
is satisfied~: the ghost equation,
peculiar of the Landau gauge~\cite{Blasi:1991xz}.
Its main consequence is a nonrenormalization theorem, which states
that, in the Landau gauge, the ghost field~$c$ enters the quantum
theory only if differentiated~($\partial_{\mu}c$). It is precisely for
the purpose of being able to take advantage of this additional
symmetry of the theory, that in~$S_k$ external sources have
been coupled not only to~$\Phi^{(k)}(\phi)$ and $\PP^{(k)}(c,\phi)$,
but also to the operators
$\partial_{a_{1}}..\partial_{a_{p-1}}\PP^{(k)}(c,\phi)$.
The ghost equation reads (see Appendix)
\eq
\FF^{a}\S=\D^{a}\ ,
\eqn{ghosteq}
where
\eq
\FF^{a} = \tint \left ( \fud{}{c^{a}} +
f^{abc}\bar{c}^{b}\fud{}{b^{c}} -
\sum_{p=1}^{2k-3} (-1)^{2k-p}
\a_{2k-p}^{a_{1}..a_{p-1}}
\fud{}{\a_{2k-p-1}^{aa_{1}..a_{p-1}}}
\right)\ ,
\eqn{a13}
and $\D^{a}$ is a classical breaking, linear in the quantum fields,
given by
\eq
\D^{a} = \tint \left (
\sum_{\varphi} f^{abc} \varphi^{*b}\varphi^{c} +
\a_{2}^{a_{1}..a_{2k-3}}
\partial_{a}\partial_{a_{1}}..\partial_{a_{2k-3}}
\PP^{(k)}
\right)\ ,
\eqn{a14}
indeed, the operator $\partial_{a}\partial_{a_{1}}..\partial_{a_{2k-3}}
\PP^{(k)}(c,\phi)$ is independent from the field~$\phi$, and it is
linear in the ghost field~$c$.

Once observed that both the ST identity~\equ{stidentity} and the ghost
equation~\equ{ghosteq}, being anomaly--free, can safely be extended to the
quantum level~\cite{Fucito:1997xm}
\eqa
\SS(\G) &=& 0 \nonumber \\
\FF^{a}\G &=& \D^{a}\ ,
\eqan{a15}
where $\G$ is the 1PI generating functional $\G=\S+O(\hbar)$, we are
able to prove our main result, $i.e.$ that the quantum insertions
$\Tr\phi^{k}\cdot\G = \Tr\phi^{k} + O(\hbar)$ have vanishing
anomalous dimensions, namely satisfy the Callan--Symanzik (CS) equation
\eq
\CC \left [ \Tr \phi^{k}(x)\cdot\G\right] =0\ ,
\eqn{result}
where $\CC$ is the CS operator
\eq
\CC \equiv \m\pad{}{\m} + \hbar \b_{g}\pad{}{g} - \hbar
\sum_{\varphi}\gamma_{\varphi}\NN_{\varphi}
\eqn{cs}
satisfying
\eq
\CC \G =0\ .
\eqn{cgamma}
In \equ{cs}, $\b_{g}$ is the $\b$--function of the gauge coupling
constant~$g$, $\g_{\varphi}$ is the anomalous dimension of the
generic field~$\varphi$, the operator~$\NN_{\varphi}$ is the counting
operator $\Tr\tint\varphi\fud{}{\varphi}$, and~$\mu$ is the
renormalization scale.

Let us take the derivative of the quantum ST identity~\equ{stidentity}
with respect to
the external source~$\a_{2k-1}$, coupled in~\equ{sk}
to~$\PP^{(k)}(c,\phi)$
\eq
\left.\fud{}{\a_{2k-1}}\SS(\G)\right|_{\a=0} =0\ ,
\eqn{derast}
where we put to zero   the set of external
sources~$\a$
appearing in~$S_k$~\equ{sk}. From the expression of the ST
operator, Eq.~\equ{derast} writes
\eq
\left. b_{\G} \fud{\G}{\a_{2k-1}}\right|_{\a=0} =
\left. \fud{\G}{\a_{0}}\right|_{\a=0}\ .
\eqn{derast1}
The operator $b_{\G}$ appearing in the above equation is the
(off--shell nilpotent) linearized quantum ST operator
\eqa
b_{\G} &=& \tint \left[
\Tr\sum_{\varphi}  \left (
\fud{\G}{\varphi^{*}} \fud{}{\varphi} +
\fud{\G}{\varphi} \fud{}{\varphi^{*}} +
\bar{c}\fud{\S}{b} \right)
\right. \\ && \left.
+ \a_{2k-1} \fud{}{\a_{0}} +
\sum_{p=2}^{2k-3} \frac{p(1-p)}{2} f^{mna_{p-1}}
\a_{2k-p-1}^{mna_{1}..a_{p-2}} \fud{}{\a_{2k-p}^{a_{1}..a_{p-1}}}
\right]\nonumber
\eqan{a16}
and it can be recognized as the quantum functional translation of the
operator~$\QQ$. The relation~\equ{derast1} is the quantum
extension of the classical one~\equ{shortvariax} between the composite
operators~$\Tr\phi^{k}$ and $\PP^{(k)}(c,\phi)$~:
\eq
D^{a_{1}.. a_{k}} \phi^{a_{1}}.. \phi^{a_{k}}\cdot\G
= b_{\G} \left(\PP^{(k)}(c,\phi)\cdot\G\right)\ .
\eqn{a17}
Owing to the ghost equation \equ{ghosteq}, the quantum insertion
$\PP^{(k)}(c,\phi)\cdot\G$ has vanishing anomalous dimension, since
the undifferentiated ghost field $c$ does not get quantum
corrections
\eq
\CC \left(\PP^{(k)}(c,\phi)\cdot\G\right) =0\ .
\eqn{pinco}
Now, since the CS operator commutes with the linearized quantum ST
operator $b_{\G}$
\eq
[\CC,b_{\G}]=0\ ,
\eqn{pallino}
from Eq.~\equ{pinco} and Eq.~\equ{pallino}
we derive our result~\equ{result}~: the
composite operators $\Tr\phi^{k}$ are finite to all orders of
perturbation theory. This result can be trivially extended  to the
class of operators
$\Tr\overline\phi^{k}$.

We would like to stress once again that the presence of matter does
not alter in any way our result~\equ{result}. Indeed,  we
needed only to consider
the symmetry transformations~\equ{qvariax} of the scalar field
$\phi$ of the $\NN=2$
gauge supermultiplet and of the Faddeev--Popov ghost field $c$. These
transformations are not altered by the presence of matter fields,
hence the matter sector is completely decoupled in our proof of
existence of protected operators. Under this respect, matter is only
an avoidable graphical charge, which we preferred to omit.
The class of protected operators discussed in this paper refers
therefore to $\NN=2$ SYM theories coupled to matter in an arbitrary
representation of the gauge group, hence to $\NN=4$ theory as well,
since for the particular choice of matter in the adjoint
representation, $\NN=4$ SYM is recovered from the $\NN=2$ case.

\section{Conclusions}

We have shown the vanishing to all orders of the perturbative expansion
of the anomalous dimension of
local single and multi-trace composite operators of scalar fields in theories
with ${\cal N}=2$ supersymmetry.
Although the proof has been given
explicitly for the pure gauge case, it holds unaltered for ${\cal N}=2$ SYM
coupled to matter belonging to  a generic representation of the gauge
group $G$, hence in particular it remains valid for the ${\cal N}=4$ case,
obtained by  putting matter multiplets in the adjoint
representation  of $G$.
Nonrenormalization properties of these operators can be better displayed
using the twisted formulation of the theory, where they have a nice
geometrical interpretation as components of the
Chern classes of a suitable universal bundle \cite{Baulieu:1988xs}.
The study of correlation functions of such operators play a relevant
role in tests of the AdS/CFT correspondence \cite{Bianchi:1999ie,
Eden:1999gh}.
In addition, it recently led to conjecture the existence of a new class
of gauge invariant operators which are protected
despite they do not obey any of the known shortening conditions
\cite{Bianchi:2001cm, Arutyunov:2001mh}.

We stress that our analysis does not rely on the superconformal
algebra. It would then be interesting to extend it to the above
mentioned operators.
Moreover, the approach we followed can be applied also to
nonconformal~${\cal N}=2$ theories which have been recently considered in
extensions of the AdS/CFT duality \cite{Polchinski:2001mx, Evans:2000ct}.
Even if a
precise map between bulk supergravity fields and boundary gauge-invariant
field theory operators, as that of the ${\cal N}=4$ superconformal theory,
is not presently known for these cases, the appearance of protected
operators in theories with a nontrivial renormalization group flow is a
very interesting feature by itself, and could motivate further analysis on
the correlation functions of these operators.

\vskip 0.5in

{\bf Acknowledgements}

We acknowledge helpful discussions with L.~Baulieu,
M.~Bianchi, S.~Kovacs, C.~Schweigert and I.M.~Singer.
A.T. is supported by a fellowship of the MURST project.
This work is partially supported by the European Commission,
RTN programme HPRN-CT-2000-00131 and by MURST.

\appendix

\section{ $\Phi^{(k)}(\phi) = \QQ\PP^{(k)}(c,\phi)$}

The fields $\phi$ and $c$ and the parameter $\om$ are assigned the
following quantum numbers
\eq
\begin{tabular}{|c|c|c||c|}
\hline
& $ \phi $ & $ c $ & $ \om $  \\ \hline
$\dim \mathrm{.}$ & $1$ & $0$ & $-1/2$ \\ \hline
$\mathcal{R}\mathrm{-charge}$ & $2$ & $0$ & $-1$  \\ \hline
$\Phi\Pi\mathrm{-charge}$ & $0$ & $1$ & $1$  \\ \hline
$\mathcal{R} + \Phi\Pi$ & $2$ & $1$ & $0$
\\ \hline
\end{tabular}
\eqn{appa1}
Therefore the most general expression for
$\Phi^{(k)}(\phi) =  D^{a_{1}.. a_{k}} \phi^{a_{1}}.. \phi^{a_{k}}$
having homogeneous $\RR+\Phi\Pi$ quantum number is the following
\eq
D^{a_{1}.. a_{k}} \phi^{a_{1}}.. \phi^{a_{k}} =
\QQ\ \left[ D^{a_{1}.. a_{k}} \sum_{p=0}^{k-1}
f_{p}^{(k)} c^{a_{1}}sc^{a_{2}}.. sc^{a_{k-p}}
\phi^{a_{k-p+1}}.. \phi^{a_{k}}\right]\ .
\eqn{appa2}
In \equ{appa2}, the coefficients $f_{p}^{(k)}$ depend on the global
ghost~$\om$ appearing in the definition of the symmetry operator
$\QQ=s+\om\d$~\equ{defQ}. The first term of the sum~\equ{appa2} is
proportional to
$D^{a_{1}.. a_{k}}
c^{a_{1}}sc^{a_{2}}.. sc^{a_{k}}$,
which is the $c^{(2k-1)}$ cocycle, belonging to the
cohomology of the BRS operator~\cite{Manes:1985df},
and known to be finite to all
orders of perturbation theory~\cite{Piguet:1992yg}. The properties of
the invariant polynomials~$\Phi^{(k)}(\phi)$ are mostly determined by
the relationship existing between them and the BRS
cocycles~$c^{(2k-1)}$.

Our first goal is to find the explicit expression for the
coefficients~$f_{p}^{(k)}$ in~\equ{appa2}. In order to obtain it,
let us observe that
\eq
\QQ sc^{a} = \om^{2} s\phi^{a}\ ,
\eqn{appa3}
and that
\eq
D^{a_{1}..a_{n}}
c^{a_{1}}sc^{a_{2}}..sc^{a_{m}}
\phi^{a_{m+1}}..\phi^{a_{n-1}}
s\phi^{a_{n}}
=
\frac{2}{n-m}
D^{a_{1}..a_{n}}
sc^{a_{1}}..sc^{a_{m}}
\phi^{a_{m+1}}..\phi^{a_{n}} \ .
\eqn{appa4}
Eq.~\equ{appa4} is a consequence of the nilpotency of the BRS
operator~$s$ and of the fact that the completely symmetric rank$-k$
tensor $D^{a_{1}..a_{k}}$ is invariant, hence it satisfies the
Jacoby--like identity
\eq
D^{p_{1}a_{2}..a_{k}}f^{p_{1}qa_{1}} +
D^{a_{1}p_{2}a_{3}..a_{k}}f^{p_{2}qa_{2}} +
\ldots +
D^{a_{1}..a_{k-1}p_{k}}f^{p_{k}qa_{k}} = 0\ .
\eqn{appa5}
Performing the $\QQ$--variation in~\equ{appa2}
and exploiting the identities (\ref{appa3}), (\ref{appa4}),
we get
\eqa
D^{a_{1}.. a_{k}} \phi^{a_{1}}.. \phi^{a_{k}}
&=&
-
\sum_{p=1}^{k-1} D^{a_{1}.. a_{k}}f^{(k)}_{p}
sc^{a_{1}}..sc^{a_{k-p}}
\phi^{a_{k-p+1}}..\phi^{a_{k}}
\nonumber\\
&& +
\sum_{p=0}^{k-2} D^{a_{1}.. a_{k}}f^{(k)}_{p}
\om^{2}
\left(1-\frac{2k}{p+1}\right)
sc^{a_{1}}..sc^{a_{k-p-1}}
\phi^{a_{k-p}}..\phi^{a_{k}}
\nonumber\\
&&
-f^{(k)}_{k-1} \om^{2} D^{a_{1}.. a_{k}} \phi^{a_{1}}.. \phi^{a_{k}}
   \nonumber   \eqan{qw}

Hence, reorganizing terms, we have
\eqa
D^{a_{1}.. a_{k}} \phi^{a_{1}}.. \phi^{a_{k}} &=&
-f^{(k)}_{k-1} \om^{2} D^{a_{1}.. a_{k}} \phi^{a_{1}}.. \phi^{a_{k}}
\label{appa7}\\
&&
+
\sum_{p=1}^{k-1}
D^{a_{1}.. a_{k}}
sc^{a_{1}}..sc^{a_{k-p}}
\phi^{a_{k-p+1}}..\phi^{a_{k}}
\left [
\om^{2}\left (
1-\frac{2k}{p}
\right )
f^{(k)}_{p-1}
-
f^{(k)}_{p}
\right ].  \nonumber
\eqan{kc}
We must therefore impose
\eqa
f^{(k)}_{k-1} &=& - \frac{1}{\om^{2}} \label{appa8}\\
f^{(k)}_{p} &=& \om^{2}\left (
                   1-\frac{2k}{p}
                        \right )
                          f^{(k)}_{p-1}\ \ 1\leq p \leq k-1\ .
			\label{appa9}
\eqan{oa}
Eqs \equ{appa8} and \equ{appa9} can be solved by a closed formula
\eq
f_{p}^{(k)} =
\frac{(-1)^{k-p}}{\om^{2(k-p)}}
\frac{(k-1)!k!}{p!(2k-p-1)!}\ \ ,\ \ 0\leq p\leq k-1\ ,
\eqn{appa10}
as can be verified by induction. This proves the relations~\equ{variax}
and~\equ{a5}.

For future reference, it may be useful to write~\equ{variax} for the
lowest values~\mbox{$k=2,3,4$}.

For $k=2$, the invariant tensor~$D^{ab}$ is just the Kronecker
delta~$\d^{ab}$, and from~\equ{variax} and~\equ{a5}
we immediately get
\eqa
\phi^{a}\phi^{a} &=&
\QQ \left (f^{(2)}_{0}c^{a}sc^{a} + f^{(2)}_{1} c^{a}\phi^{a}
\right) \nonumber \\
&=& \QQ \left (
\frac{1}{3\om^{4}} c^{a}sc^{a} - \frac{1}{\om^{2}} c^{a}\phi^{a}
\label{appa11}\right )\ .
\eqan{bb}

For $k=3$, there is only one completely symmetric tensor
$D^{abc}=d^{abc}$, and
\eqa
d^{abc}\phi^{a}\phi^{b}\phi^{c} &=&
\QQ d^{abc}\left (
f^{(3)}_{0}c^{a}sc^{b}sc^{c} +
f^{(3)}_{1}c^{a}sc^{b}\phi^{c} +
f^{(3)}_{2}c^{a}\phi^{b}\phi^{c}
\right) \\
&=&
\QQ d^{abc}\left (
-\frac{1}{10\om^{6}} c^{a}sc^{b}sc^{c}
+\frac{1}{2\om^{4}} c^{a}sc^{b}\phi^{c}
-\frac{1}{\om^{2}} c^{a}\phi^{b}\phi^{c}
\right)\ .\nonumber
\eqan{appa12}

For $k=4$, there are more than one symmetric tensors~$D^{abcd}$
\eqa
D^{abcd}\phi^{a}\phi^{b}\phi^{c}\phi^{d} &=&
\QQ D^{abcd} \left(
f^{(4)}_{0}c^{a}sc^{b}sc^{c}sc^{d} +
f^{(4)}_{1}c^{a}sc^{b}sc^{c}\phi^{d} \right.
\nonumber \\
&&\quad +\left.
f^{(4)}_{2}c^{a}sc^{b}\phi^{c}\phi^{d} +
f^{(4)}_{3}c^{a}\phi^{b}\phi^{c}\phi^{d}
\right) \\
&=&
\QQ D^{abcd} \left(
\frac{1}{35\om^{8}} c^{a}sc^{b}sc^{c}sc^{d} -
\frac{1}{5\om^{6}} c^{a}sc^{b}sc^{c}\phi^{d} \right.
\nonumber \\
&&\quad +\left.
\frac{3}{5\om^{4}} c^{a}sc^{b}\phi^{c}\phi^{d} -
\frac{1}{\om^{2}}c^{a}\phi^{b}\phi^{c}\phi^{d}
\right)\ .\nonumber
\eqan{appa13}

\section{The Slavnov--Taylor identity}

Our task is to extend the action of $\QQ$ on the set of external
$\a$--sources introduced in
\eq
S_k = \om^{2k} \tint \left (
\a_{0} \Phi^{(k)} +
\a_{2k-1} \PP^{(k)} +
\sum_{p=2}^{2k-2} \a^{a_{1}..a_{p-1}}_{2k-p}
\partial_{a_{1}}..\partial_{a_{p-1}}\PP^{(k)}
\right)\ ,
\eqn{appb1}
by demanding
\eq
\QQ\ S_k = 0\ .
\eqn{appb2}
We have
\eqa
\QQ S_k &=& \om^{2k} \tint \left \{
(\QQ\a_{0})\Phi^{(k)}
\right.
\nonumber\\ &&
+ \sum_{p=1}^{2k-2} \left [
(\QQ\a_{2k-p}^{a_{1}..a_{p-1}})
\partial_{a_{1}}..\partial_{a_{p-1}}  \PP^{(k)}  \label{appb3}\right. \\
&& \quad \left.
+(-1)^{2k-p} \a_{2k-p}^{a_{1}..a_{p-1}}
(\QQ \partial_{a_{1}}..\partial_{a_{p-1}} \PP^{(k)} )\left.
\right]\right\}\ ,\nonumber
\eqan{aappb3}
where we taked into account the $\QQ$--invariance of the
polynomials~$\Phi^{(k)}(\phi)$ and the statistics of
the~$\a$--sources, given by~$(-1)^{2k-p}$.

In order to evaluate~\equ{appb3}, we observe that the anticommutator
between the derivative with respect to the ghost field~$c$
($\partial_{a}$) and the symmetry operator~$\QQ$, gives the generators
of the rigid gauge transformations~$\HH^{a}$
\eq
\left\{\partial_{a},\QQ\right\}X^{b} = -\HH^{a}X^{b} = - f^{abc} X^{c}\ ,
\eqn{appb4}
where $X^{a}$ is a generic function of the fields $\phi$ and $c$
and belongs to the adjoint representation of the gauge group. From
the obvious relations
\eq
\HH^{a}\Phi^{(k)}(\phi)=\HH^{a}\PP^{(k)}(c,\phi)=
\partial_{a}\QQ\PP^{(k)}(c,\phi)=0\ ,
\eqn{appb5}
we derive
\eq
\QQ\partial_{a}\PP^{(k)} = 0
\eqn{aappb5}
and, in general,
\eq
\alpha^{a_1..a_m}\QQ \partial_{a_{1}}..\partial_{a_{m}} \PP^{(k)} =
         (-1)^{m+1}\ \frac{m(m-1)}{2}
         f^{a_{1}a_{2}p}\partial_{a_{3}}..
              \partial_{a_{m}}\partial_{p}\PP^{(k)}\alpha^{a_1..a_m}\ .
	    \eqn{appb10}
Hence \equ{appb3} writes
\eqa
\QQ S_k &=& \om^{2k} \tint \left \{
(\QQ\a_{0})\Phi^{(k)}
-\a_{2k-1}\Phi^{(k)}
\right.
\nonumber \\ &&
+ \sum_{p=1}^{2k-2} \left [
(\QQ\a_{2k-p}^{a_{1}..a_{p-1}})
\partial_{a1}..\partial_{a_{p-1}}  \PP^{(k)} \nonumber\right. \\
&& \quad \left.
+\frac{(p-1)(p-2)}{2}\a_{2k-p}^{a_{1}..a_{p-1}} f^{a_{1}a_{2}q}
\partial_{a_{3}}..\partial_{a_{p-1}}\partial_{a_{q}}\PP^{(k)}
\left.\right]\right\}\  \label{appb11}\\
&=&
\om^{2k} \tint \left \{
(\QQ\a_{0})\Phi^{(k)}
-\a_{2k-1}\Phi^{(k)}
+ (\QQ\a_{2k-1})\PP^{(k)}
\right. \nonumber \\
&&\quad \left.
+
(\QQ\a_{2}^{a_{1}..a_{2k-3}})\partial_{a_{1}}..\partial_{a_{2k-3}}\PP^{(k)}
\right.
\nonumber \\ &&  \left.
+ \sum_{p=2}^{2k-3} \left [
(\QQ\a_{2k-p}^{a_{1}..a_{p-1}})
+\frac{p(p-1)}{2} \a_{2k-p-1}^{rsa_{1}..a_{p-2}} f^{rsa_{p-1}}
\right ]
\partial_{a_{1}}..\partial_{a_{p-1}}\PP^{(k)}
\right\}
\ .\nonumber \\
\nonumber\eqan{aappb7}
In order that the whole expression~$\QQ S_k$ vanish, by identifying
terms containing equal number of ghost fields~$c$, we have
\eqa
\QQ\a_{0} &=& \a_{2k-1} \\
\QQ\a_{2}^{a_{1}..a_{2k-3}} &=& 0 \\
\QQ\a_{2k-1} &=& 0 \ ,
\eqan{rr}
and
\eq
\QQ\a_{2k-p}^{a_{1}..a_{p-1}} =
- \frac{p(p-1)}{2} f^{rsa_{p-1}} \a_{2k-p-1}^{rsa_{1}..a_{p-2}}\ \
2\leq p\leq 2k-3\ .
\eqn{kk}
Hence, the implementation of the ST identity to the $\a$--sector is
given by
\eq
\SS^{(k)}\S = \tint \left (
\a_{2k-1} \fud{}{\a_{0}} +
\sum_{p=2}^{2k-3} \frac{p(1-p)}{2} f^{mna_{p-1}}
\a_{2k-p-1}^{mna_{1}..a_{p-2}} \fud{}{\a_{2k-p}^{a_{1}..a_{p-1}}}
\right) \S\ ,
\eqn{apppb}
which coincides with \equ{a12}.

\section{The Ghost Equation}

To extend the ghost equation, let us take the functional
derivative with respect to ghost field~$c$ of the action
functional~$S_k$
\eqa
\d_{a} S_{k} &=& \sum_{p=1}^{2k-2} (-1)^{2k-p}
\a_{2k-p}^{a_{1}..a_{p-1}}
\partial_{a}\partial_{a_{1}}..\partial_{a_{p-1}}\PP^{(k)} \label{appc1} \\
&=&
\sum_{p=1}^{2k-3} (-1)^{2k-p}
\a_{2k-p}^{a_{1}..a_{p-1}} \fud{\S}{\a_{2k-p-1}^{aa_{1}..a_{p-1}}}
+\tint\ \a_{2}^{a_{1}..a_{2k-3}}
\partial_{a}\partial_{a_{1}}..\partial_{a_{2k-3}}\PP^{(k)} \nonumber
\eqan{as}
Notice
that, according to the expression~\equ{variax}, the maximum number of
ghosts~$c$ present in~$\PP^{(k)}(c,\phi)$ is~$2k-1$,
in the term~$c^{(2k-1)}$, whereas the
maximum number of its ghost derivatives in the last term at the r.h.s
of~\equ{appc1} is~$2k-2$, hence this term is linear in~$c$ and independent
from~$\phi$, $i.e.$ it represents a classical breaking.

Therefore, the complete ghost equation reads
\eq
\FF^{a}\S=\D^{a}\ ,
\eqn{ap}
where
\eq
\FF^{a} = \tint \left ( \fud{}{c^{a}} +
f^{abc}\bar{c}^{b}\fud{}{b^{c}} -
\sum_{p=1}^{2k-3} (-1)^{2k-p}
\a_{2k-p}^{a_{1}..a_{p-1}}
\fud{}{\a_{2k-p-1}^{aa_{1}..a_{p-1}}}
\right)\ ,
\eqn{appbb}
and $\D^{a}$ is the classical breaking
given by
\eq
\D^{a} = \tint \left (
\sum_{\varphi} f^{abc} \varphi^{*b}\varphi^{c} +
\a_{2}^{a_{1}..a_{2k-3}}
\partial_{a}\partial_{a_{1}}..\partial_{a_{2k-3}}
\PP^{(k)}
\right)\ .
\eqn{appkpb}

}
\end{document}